\documentclass[%
 reprint,
superscriptaddress,
 amsmath,amssymb,
 aps,
]{revtex4-2}

\usepackage{graphicx}% Include figure files
\usepackage{dcolumn}% Align table columns on decimal point
\usepackage{bm}% bold math
\usepackage{upgreek}

\begin{document}

\title{On-chip control of the coherence matrix of four-mode partially coherent light: rank, entropy, and modal Stokes parameters}

\author{Amin Hashemi}
\affiliation{CREOL, The College of Optics \& Photonics, University of Central Florida, Orlando, FL 32816, USA}
\affiliation{These authors contributed equally to this work}

\author{Abbas Shiri}
\affiliation{CREOL, The College of Optics \& Photonics, University of Central Florida, Orlando, FL 32816, USA}
\affiliation{These authors contributed equally to this work}

\author{Bahaa E. A. Saleh}
\affiliation{CREOL, The College of Optics \& Photonics, University of Central Florida, Orlando, FL 32816, USA}

\author{Andrea Blanco-Redondo}
\affiliation{CREOL, The College of Optics \& Photonics, University of Central Florida, Orlando, FL 32816, USA}

\author{Ayman F. Abouraddy}
\affiliation{CREOL, The College of Optics \& Photonics, University of Central Florida, Orlando, FL 32816, USA}
\affiliation{raddy@creol.ucf.edu}

%% To be edited by editor
% \dates{Compiled \today}

%\ociscodes{(140.3490) Lasers, distributed feedback; (060.2420) Fibers, polarization-maintaining;(060.3735) Fiber Bragg gratings.}

%% To be edited by editor
% \doi{\url{http://dx.doi.org/10.1364/XX.XX.XXXXXX}}

\begin{abstract}
Partially coherent light offers salutary capabilities in optical information processing that cannot be matched by coherent light. To date, this `coherence advantage' has been confirmed in proof-of-principle optical communications protocols using bulk optics. Taking full advantage of such opportunities necessitates processing multimode partially coherent light in integrated photonics platforms that alone provide the requisite stability for cascaded operations on a large scale. Here we demonstrate on-chip manipulation of four-mode partially coherent light described by a $4\times4$ Hermitian coherence matrix. Starting with generic maximally incoherent light, we utilize an on-chip hexagonal mesh of Mach-Zehnder interferometers to perform all the unitary and non-unitary tasks that are critical for realizing structured coherence: controlling the coherence rank (the number of non-zero eigenvalues of the coherence matrix); tuning the field entropy; molding the structure of the coherence matrix via $4\times4$ unitary transformations constructed out of sequences of $2\times2$ unitaries acting on pairs of modes; and tomographic reconstruction of the coherence matrix by measuring the modal Stokes parameters associated with Kronecker-Pauli matrices. These results confirm the scalability of utilizing $2\times2$ on-chip building blocks for the synthesis and reconstruction of high-dimensional coherence matrices, and provide a decisive step towards large-scale on-chip manipulation of massively moded partially coherent light for applications in optical information processing.  
\end{abstract}

\maketitle
\section{Introduction}

Because all natural sources of light -- from solar and stellar radiation to fluorescence and luminescence -- are partially coherent, the  study of optical coherence has long been pursued \cite{Michelson1890PM,Michelson1891PM2,Zernicke} and is now a well-established branch of optical physics \cite{Mandel65RMP}. Since its inception \cite{Wolf54NCC,Wolf55PRSL,Wolf59INC,Karczewski63INC,Perina72Book}, the spatiotemporal dynamics of partially coherent light has been described within a framework of continuous correlation functions (in addition to polarization) \cite{Wolf07Book,Agarwal20PO}, with applications extending from propagation in turbulent media \cite{Gbur02JOSAA,Ponomarenko02OL,Dogariu03OL}, to sensing and metrology \cite{Baleine04JOSAA,Redding12NP}. However, recent findings making use of a discrete formulation of optical coherence have unveiled a `coherence advantage' in optical information processing, wherein partially coherent light outperforms its coherent counterpart \cite{Nardi22OL,Dong24Nature,Harling25APLP}. One example occurs in communicating optically across a channel supporting $N$ modes (such as a multimode fiber or a bundle of single-mode fibers), where coherent light supports $\mathcal{O}(N)$ independent communications signals, whereas partially coherent light supports $\mathcal{O}(N^{2})$ \cite{Nardi22OL}. In another example, the coherence rank (the number of non-zero eigenvalues of the coherence matrix) was exploited as a scattering-immune information carrier -- even when the scattering is strong (arbitrary modal coupling) and the channel characteristics change from bit to bit (thus precluding the use of adaptive optics) \cite{Harling25APLP}.

These applications, along with other recent results \cite{Dong24Nature,Miller25Optica}, motivate the transition from free-space optics -- which remains to date the exclusive domain for the experimental investigation of partially coherent light -- to integrated-photonics platforms for the manipulation of spatial coherence in massive arrays of single-mode waveguides. Such platforms \cite{Shen17NP,Bogaerts20Nature,Capmany20Book,Ashtiani22Nature} alone offer the requisite conditions for versatile exploitation of the `coherence advantage' in optical information processing: small footprints that enable scaling up the number of optical modes, high-speed processing, and the interferometric stability needed for large-scale cascading of optical transformations. Substantial progress has been made in on-chip processing of quantum and classical optical fields, which has nevertheless focused on pure states in the former and coherent light in the latter. It remains now for these advances to be harnessed for the processing of partially coherent light. On-chip manipulation of two-mode partially coherent optical fields has been recently demonstrated experimentally, including molding the two-mode coherence matrix via $2\times2$ unitary transformations (`unitaries' henceforth for brevity) and tuning the degree of coherence via non-unitaries \cite{Hashemi26arxiv}. This is a critical step towards the construction of larger-dimensional unitaries \cite{Reck94PRL,Miller17OE,Bogaerts20Nature,Saleh25book}, for which $2\times2$ unitaries are the essential building blocks. 

Here we demonstrate for the first time scalable on-chip coupling, unitary and non-unitary manipulation, and reconstruction of partial coherence in the spatial domain, realized here using four-mode partially coherent optical fields. The integrated-photonics platform utilized here comprises a large hexagonal array of programmable Mach-Zehnder interferometers (MZIs). We first produce a generic four-mode mutually incoherent spatial field in single-mode fibers starting from a short-coherence-length laser diode that can be efficiently coupled to on-chip single-mode waveguides. Second we control the coherence rank of the field by selectively eliminating on chip the modal weights (a non-unitary linear operation), from rank-1 fields (fully coherent light extracted from the initial generic incoherent light) to maximally incoherent rank-4 fields.  Third, we tune the entropy of the field without changing the rank, another non-unitary linear operation. We further synthesize iso-entropy rank-3 and rank-4 optical fields that cannot be inter-converted unitarily \cite{Harling24PRA2}. Fourth, we construct examples of on-chip $4\times4$ unitaries that modify the structure of the coherence matrix, thereby enabling control over the real and imaginary parts of its off-diagonal elements. Fifth, we reconstruct the coherence matrix by measuring the modal Stokes parameters mediated by $4\times4$ Kronecker-Pauli matrices formed of direct products of the usual $2\times2$ Pauli matrices. These results demonstrate that on-chip platforms are capable of performing all the critical tasks needed for the reliable processing of multimoded partially coherent light, which is expected to enable new fundamental studies of partial coherence \cite{Waller12NP,Okoro17Optica,Harling24PRA} and utilizing such optical fields in communications, cryptography, spectroscopy, sensing and metrology, and computation.

\section{Coherence matrix for a four-mode field}

We consider an optical field spanned by four modes labeled $\{|1\rangle,|2\rangle,|3\rangle,|4\rangle\}$ in the Dirac notation, which correspond to the fields in four single-mode on-chip waveguides. A coherent field in this basis is described by the vector $|E\rangle=\sum_{j=1}^{4}E_{j}|j\rangle$, where $E_{j}$ are the modal amplitudes, and the normalization $\langle E|E\rangle=1$ entails that $\sum_{j=1}^{4}|E_{j}|^{2}=1$. We refer to the squared magnitude $|E_{j}|^{2}$ as the modal weight, which is the fraction of power associated with the $j^{\mathrm{th}}$-mode. A partially coherent field is described by the $4\times4$ coherence matrix \cite{Gamo64PO,Gori06OL,Abouraddy17OE,Kagalwala13NP,Okoro17Optica,Harling22OE,Harling23JO}:
\begin{equation}
\mathbf{G}=\left(\!\begin{array}{cccc}
G_{11}&G_{12}&G_{13}&G_{14}\\
G_{21}&G_{22}&G_{23}&G_{24}\\
G_{31}&G_{32}&G_{33}&G_{34}\\
G_{41}&G_{42}&G_{43}&G_{44}
\end{array}\!\right),
\end{equation}
where $G_{jk}=\langle E_{j}E_{k}^{*}\rangle$, $j,k=1,2,3,4$, and $\langle\cdot\rangle$ is a statistical average over an ensemble. The coherence matrix is Hermitian $\mathbf{G}^{\dagger}=\mathbf{G}$, so that $G_{jk}=G_{kj}^{*}$, and $\mathbf{G}$ is normalized so that $\mathrm{Tr}\{\mathbf{G}\}=1$, where $\mathrm{Tr}\{\cdot\}$ refers to the matrix trace. The diagonal elements $G_{jj}$ are the modal weights, and the off-diagonal element $G_{jk}$, $j\neq k$, determines the interference visibility resulting from superposing the $j^{\mathrm{th}}$ and $k^{\mathrm{th}}$ modes. 

The Hermitian coherence matrix $\mathbf{G}$ can be diagonalized via a $4\times4$ unitary $\hat{U}$ such that $\hat{U}\mathbf{G}\hat{U}^{\dagger}=\mathbf{G}^{\mathrm{D}}$, where $\mathbf{G}^{\mathrm{D}}$ is a diagonal coherence matrix,
\begin{equation}
\mathbf{G}^{\mathrm{D}}=\left(\!\begin{array}{cccc}
\lambda_{1}&0&0&0\\
0&\lambda_{2}&0&0\\
0&0&\lambda_{3}&0\\
0&0&0&\lambda_{4}
\end{array}\!\right)=\mathrm{diag}\{\lambda_{1},\lambda_{2},\lambda_{3},\lambda_{4}\},
\end{equation}
$\lambda_{1},\lambda_{2},\lambda_{3}$, and $\lambda_{4}$ ($\lambda_{j}\geq0$) are the eigenvalues of $\mathbf{G}$ arranged in descending order, $\sum_{j=1}^{4}\lambda_{j}=1$, and the shorthand $\mathrm{diag}\{\cdots\}$ indicates a diagonal matrix with the entries representing the diagonal elements. The entropy of a coherence matrix is \cite{Gamo64PO}:
\begin{equation}\label{eq:Entropy}
S=-\mathrm{Tr}\{\mathbf{G}\log_{2}\mathbf{G}\}=-\sum_{j=1}^{4}\lambda_{j}\log_{2}\lambda_{j},
\end{equation}
with $0\leq S\leq2$. A \textit{diagonal} coherence matrix corresponds to a configuration in which the correlations between all pairs of modes have been eliminated (no interference fringes are observed when they are overlapped).

We make use of a fourfold classification of partially coherent four-mode fields according to the `coherence rank', which is defined as the number of non-zero eigenvalues of $\mathbf{G}$ \cite{Harling24PRA,Harling24PRA2}. Rank-1 fields ($\lambda_{2}=\lambda_{3}=\lambda_{4}=0$) are coherent with $S=0$, corresponding to a complete lack of randomness. We take $\mathbf{G}_{1}=\mathrm{diag}\{1,0,0,0\}$ as a representative rank-1 coherence matrix. Rank-2 fields ($\lambda_{3}=\lambda_{4}=0$) have $0<S\leq1$~bit, and we take the maximum-entropy coherence matrix $\mathbf{G}_{2}=\mathrm{diag}\{\tfrac{1}{2},\tfrac{1}{2},0,0\}$ ($S=1$~bit) as representative. Rank-3 fields ($\lambda_{4}=0$) have $0<S\leq\log_{2}3$~bits, and we take the maximum-entropy coherence matrix $\mathbf{G}_{3}=\mathrm{diag}\{\tfrac{1}{3},\tfrac{1}{3},\tfrac{1}{3},0\}$ ($S=\log_{2}3\approx1.585$~bits) as representative. Finally, all the eigenvalues of rank-4 fields are non-zero, with $0<S\leq2$~bits, and we take the maximum-entropy coherence matrix $\mathbf{G}_{4}=\mathrm{diag}\{\tfrac{1}{4},\tfrac{1}{4},\tfrac{1}{4},\tfrac{1}{4}\}$ ($S=2$~bits) as representative. These four representative coherence matrices, $\mathbf{G}_{1}$ through $\mathbf{G}_{4}$, are the basis for the coherence-rank-communications experiment in Ref.~\cite{Harling25APLP}, and we thus take them as starting points for the unitaries to be implemented here.

\section{On-chip construction of a general unitary}

We make use in our experiments of the hexagonal mesh of 72~MZIs (iPronics Smartlight Processor) depicted in Fig.~\ref{fig:2X2Unitaries}(a). We combine MZIs to construct general $2\times2$ unitaries operating on two modes at a time, a sequence of which can be assembled into a $4\times4$ unitary.

\subsection{Two-mode on-chip unitaries}

\begin{figure}[t!]
\centering
\includegraphics[width=8.4cm]{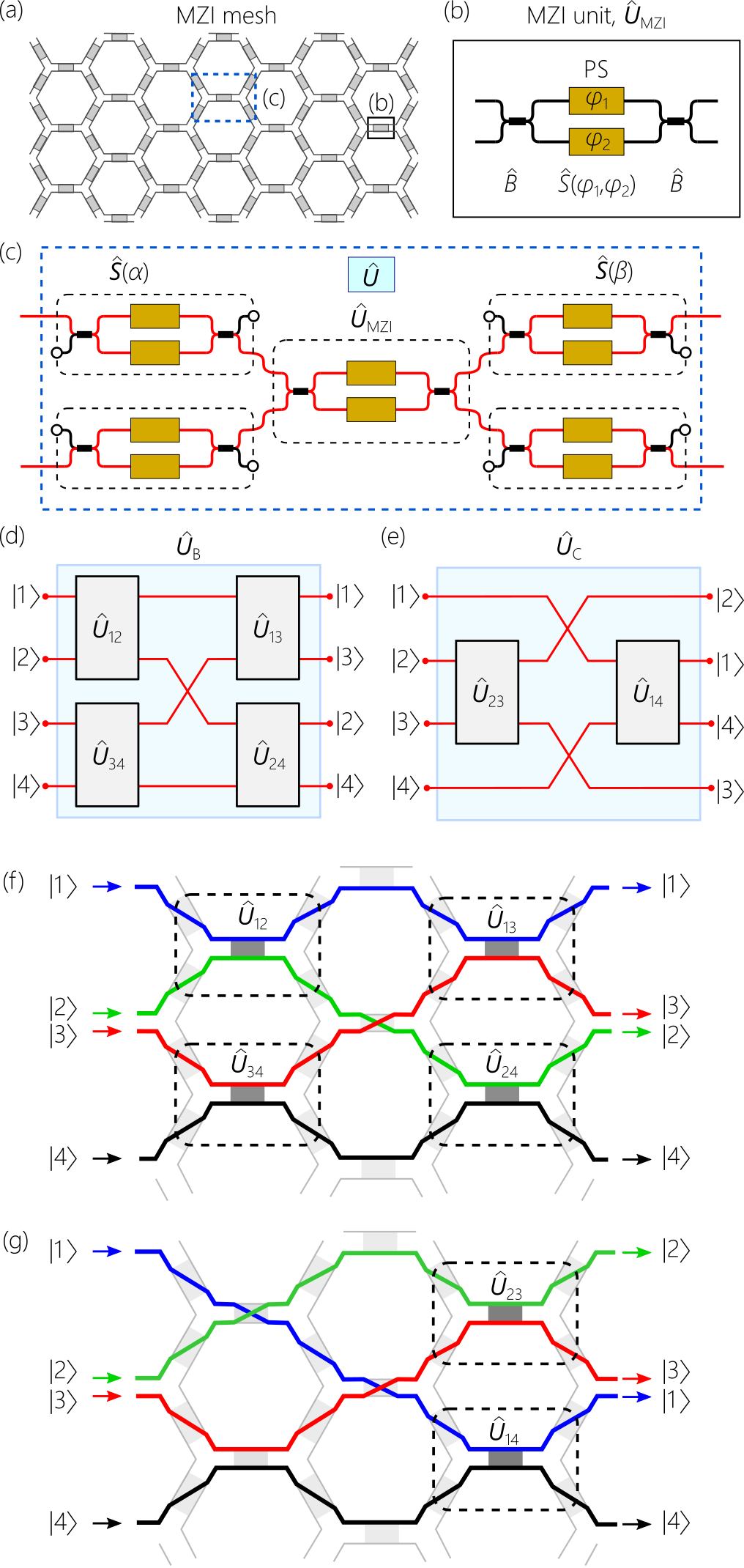} %{Fig1.jpg}
\caption{(a) Chip layout of the MZI mesh. Each colored rectangle is an MZI, and lines are single-mode waveguides. (b) Structure of a single MZI (Eq.~\ref{eq:MZI}). PS: Phase shifter. (c) A general $2\times2$ unitary (Eq.~\ref{eq:GeneralUnitary}) formed of the MZIs highlighted in (a). (d) Construction of the unitary $\hat{U}_{B}$ and (e) $\hat{U}_{C}$ out of sequences of $2\times2$ unitaries. (f) Chip layouts for $\hat{U}_{B}$ and (g) for $\hat{U}_{C}$.}
\label{fig:2X2Unitaries}
\end{figure}

We take the building-block $2\times2$ unitaries to be in the form:
\begin{equation}\label{eq:GeneralUnitary}
\hat{U}(\delta,\alpha)=ie^{i\varphi}\left(\begin{array}{cc}\sin\tfrac{\delta}{2}&e^{-i\alpha}\cos\tfrac{\delta}{2}\\e^{i\alpha}\cos\tfrac{\delta}{2}&-\sin\tfrac{\delta}{2}\end{array}\right).
\end{equation}
Each on-chip MZI is formed of a phase operator $\hat{S}(\varphi_{1},\varphi_{2})=\left(\begin{array}{cc}e^{i\varphi_{1}}&0\\0&e^{i\varphi_{2}}\end{array}\right)$ sandwiched between two symmetrical couplers described by the beam-splitter operator $\hat{B}=\tfrac{1}{\sqrt{2}}\left(\begin{array}{cc}1&i\\i&1\end{array}\right)$ [Fig.~\ref{fig:2X2Unitaries}(b)], resulting in the following unitary:
\begin{equation}\label{eq:MZI}
\hat{U}_{\mathrm{MZI}}(\delta,\varphi)=\hat{B}\hat{S}(\varphi_{1},\varphi_{2})\hat{B}=ie^{i\varphi}\left(\begin{array}{cc}\sin\tfrac{\delta}{2}&\cos\tfrac{\delta}{2}\\\cos\tfrac{\delta}{2}&-\sin\tfrac{\delta}{2}\end{array}\right),    
\end{equation}
where $\varphi=\tfrac{1}{2}(\varphi_{1}+\varphi_{2})$ and $\delta=\varphi_{1}-\varphi_{2}$. Sandwiching this MZI between phase operators $\hat{S}(\alpha)=\mathrm{diag}\{e^{i\alpha/2},e^{-i\alpha/2}\}$ and $\hat{S}(\beta)=\mathrm{diag}\{e^{i\beta/2},e^{-i\beta/2}\}$ -- each implemented, in turn, by MZIs [Fig.~\ref{fig:2X2Unitaries}(c)]. These introduce phases $\alpha$ and $\beta$ between the two modes, respectively, thereby yielding the general unitary in Eq.~\ref{eq:GeneralUnitary} after setting $\beta=-\alpha$: $\hat{U}(\delta,\alpha)=\hat{S}(-\alpha)\hat{U}_{\mathrm{MZI}}(\delta,\varphi)\hat{S}(\alpha)$.

\subsection{Four-mode on-chip unitaries}

Using such $2\times2$ unitaries, in addition to phases added to each mode, an arbitrary $4\times4$ unitary can be constructed as established in quantum information processing \cite{Reck94PRL,Zukowski97PRA}. We realize in our experiments here three distinct unitaries. The first is the identity $\hat{U}_{A}=\hat{\mathcal{I}}_{4}$, where $\hat{\mathcal{I}}_{4}$ is the $4\times4$ identity matrix, which allows us to validate the process of coherence-matrix reconstruction. The other unitaries, $\hat{U}_{\mathrm{B}}$ and $\hat{U}_{\mathrm{C}}$, are illustrated schematically in Fig.~\ref{fig:2X2Unitaries}(d,e), where:
\begin{eqnarray}\label{eq:U_BC}
\hat{U}_{\mathrm{B}}&=&
\frac{1}{2}\left(\begin{array}{cccc}
1&1&-i&i\\1&-1&-i&-i\\i&i&-1&1\\-i&i&1&1
\end{array}\right),\nonumber\\
\hat{U}_{\mathrm{C}}&=&\frac{1}{2}\left(\!\begin{array}{cccc}1&0&0&\sqrt{3}e^{-i\pi/3}\\0&\sqrt{3}&e^{-i\pi/6}&0\\0&e^{i\pi/6}&-\sqrt{3}&0\\\sqrt{3}e^{i\pi/3}&0&0&-1\end{array}\!\right).
\end{eqnarray}
These two unitaries are constructed from the $2\times2$ unitary building blocks [Fig.~\ref{fig:2X2Unitaries}(c)], each operating on the pair of modes identified by their indices, given by:
\begin{eqnarray}
\hat{U}_{12}&=&\tfrac{1}{\sqrt{2}}\left(\begin{array}{cc}1&1\\1&-1\end{array}\right),U_{34}=\tfrac{1}{\sqrt{2}}\left(\begin{array}{cc}1&-1\\-1&-1\end{array}\right),\nonumber\\
\hat{U}_{13}&=&\tfrac{1}{\sqrt{2}}\left(\begin{array}{cc}1&-i\\i&-1\end{array}\right),\hat{U}_{24}=\tfrac{1}{\sqrt{2}}\left(\begin{array}{cc}1&i\\-i&-1\end{array}\right),\nonumber\\
\hat{U}_{23}&=&\tfrac{1}{2}\left(\!\begin{array}{cc}\sqrt{3}&e^{-i\pi/6}\\e^{i\pi/6}&-\sqrt{3}\end{array}\!\right),U_{14}=\tfrac{1}{2}\left(\!\begin{array}{cc}1&\sqrt{3}e^{-i\pi/3}\\\sqrt{3}e^{i\pi/3}&-1\end{array}\!\right).
\end{eqnarray}
These $2\times2$ unitaries are obtained from Eq.~\ref{eq:GeneralUnitary} by appropriate settings for $\delta$ and $\alpha$. The most general $4\times4$ unitary is formed of a concatenation of unitaries in the form of $\hat{U}_{\mathrm{B}}$ and $\hat{U}_{\mathrm{C}}$ \cite{Reck94PRL}. We depict in Fig.~\ref{fig:2X2Unitaries}(f,g) the chip layouts for realizing $\hat{U}_{\mathrm{B}}$ and $\hat{U}_{\mathrm{C}}$, where we identify the modal pathways with different colors. 

%$\hat{U}_{12}$ by setting $\delta=\tfrac{\pi}{2}$ and $\alpha=0$; $\hat{U}_{34}$, $\delta=\tfrac{\pi}{2}$ and $\alpha=\pi$; $\hat{U}_{13}$, $\delta=\tfrac{\pi}{2}$ and $\alpha=\tfrac{\pi}{2}$; $\hat{U}_{24}$, $\tfrac{\pi}{2}$ and $\alpha=-\tfrac{\pi}{2}$; $\hat{U}_{23}$, $\delta=\tfrac{2\pi}{3}$ and $\alpha=\tfrac{\pi}{6}$; and $\hat{U}_{14}$, $\delta=\tfrac{\pi}{3}$ and $\alpha=\tfrac{\pi}{3}$

\begin{figure*}[t!]
\centering
\includegraphics[width=13.2cm]{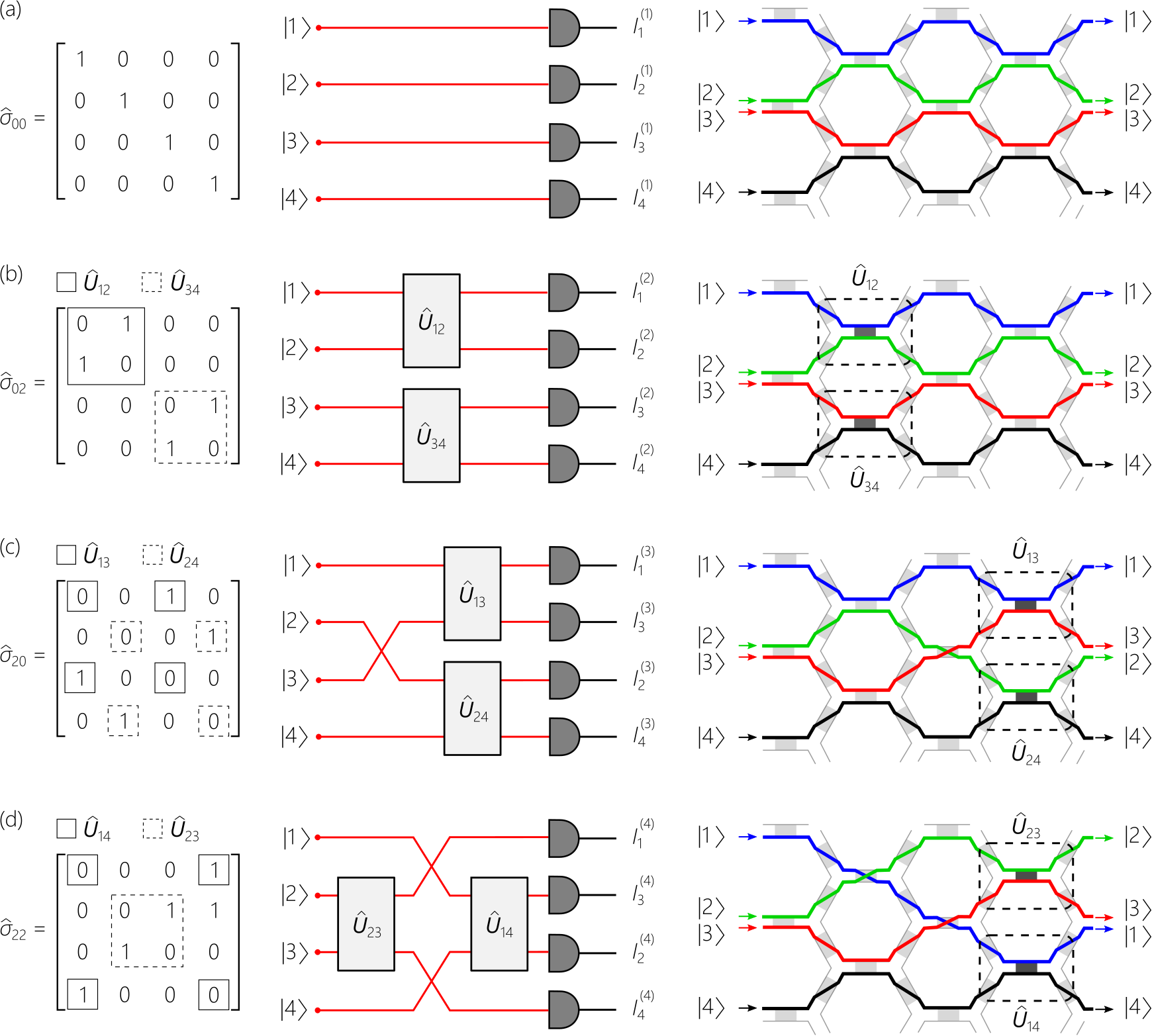} %{Fig1.jpg}
\caption{Measuring the modal Stokes parameters from which the $4\times4$ coherence matrix $\mathbf{G}$ is reconstructed. The left column highlights the structure of the relevant $4\times4$ Kronecker-Pauli matrix; the middle shows the conceptual measurement scheme; and the right depicts the chip layout. (a) Measuring $s_{00}$, $s_{01}$, $s_{10}$, and $s_{11}$. The Kronecker-Pauli matrices $\hat{\sigma}_{00}$, $\hat{\sigma}_{01}$, $\hat{\sigma}_{10}$, and $\hat{\sigma}_{11}$ are diagonal ($\hat{\sigma}_{00}=\hat{\sigma}_{0}\otimes\hat{\sigma}_{0}$ shown). Detectors record $I_{1}^{(1)}$, $I_{2}^{(1)}$, $I_{3}^{(1)}$, and $I_{4}^{(1)}$, and the modal Stokes parameters are obtained using Eq.~\ref{eq:S_00_01_10_11}. (b) Measuring $s_{02}$, $s_{03}$, $s_{12}$, and $s_{13}$. The Kronecker-Pauli matrices $\hat{\sigma}_{02}$, $\hat{\sigma}_{03}$, $\hat{\sigma}_{12}$, and $\hat{\sigma}_{13}$ are block diagonal ($\hat{\sigma}_{02}=\hat{\sigma}_{0}\otimes\hat{\sigma}_{2}$ shown). Modal Stokes parameters $s_{02}$ and $s_{12}$ are obtained using Eq.~\ref{eq:S02_12} after setting $\hat{U}_{12}=\hat{U}_{34}=\hat{U}_{2}$, and $s_{03}$ and $s_{13}$ using Eq.~\ref{eq:S03_13} after setting $\hat{U}_{12}=\hat{U}_{34}=\hat{U}_{3}$. (c) Measuring $s_{20}$, $s_{21}$, $s_{30}$, and $s_{31}$. Modal Stokes parameters $s_{20}$, $s_{21}$ are obtained using Eq.~\ref{eq:S20_21} after setting $\hat{U}_{13}=\hat{U}_{24}=\hat{U}_{2}$, and $s_{30}$, $s_{31}$ using Eq.~\ref{eq:S30_31} after setting $\hat{U}_{13}=\hat{U}_{24}=\hat{U}_{3}$. (d) Measuring $s_{22}$, $s_{33}$, $s_{23}$, and $s_{32}$. Modal Stokes parameters $s_{22}$ and $s_{33}$ are obtained using Eq.~\ref{eq:S22_33} after setting $\hat{U}_{23}=\hat{U}_{14}=\hat{U}_{2}$, and $s_{23}$ and $s_{32}$ using Eq.~\ref{eq:S23_32} after setting $\hat{U}_{23}=\hat{U}_{14}=\hat{U}_{3}$.}
\label{fig:StokesParameters}
\end{figure*}

\section{Reconstructing the coherence matrix}

\subsection{Modal Stokes parameters for a two-mode field}

To elucidate our approach to the reconstruction of a $4\times4$ coherence matrix representing a four-mode partially coherent field, we first establish the reconstruction of its $2\times2$ counterpart. A $2\times2$ Hermitian coherence matrix for modes $|1\rangle$ and $|2\rangle$ is expanded in terms of Pauli matrices, $\mathbf{G}=\tfrac{1}{2}\sum_{j=0}^{3}s_{j}\hat{\sigma}_{j}$, where $\hat{\sigma}_{0}=\hat{\mathcal{I}}_{2}$ is the $2\times2$ identity matrix, and the other Pauli matrices $\{\hat{\sigma}_{j}\}$ are:
\begin{equation}\label{eq:Pauli}
\hat{\sigma}_{1}=\left(\begin{array}{cc}1&0\\0&-1\end{array}\right),\;
\hat{\sigma}_{2}=\left(\begin{array}{cc}0&1\\1&0\end{array}\right),\;
\hat{\sigma}_{3}=\left(\begin{array}{cc}0&-i\\i&0\end{array}\right),
\end{equation}
so that the real expansion coefficients $s_{j}=\mathrm{Tr}\{\hat{\sigma}_{j}\mathbf{G}\}$ are the modal Stokes parameters, which are obtained by measuring the modal weights $I_{1}=G_{11}$ and $I_{2}=G_{22}$ (the fractions of power associated with modes $|1\rangle$ and $|2\rangle$) after traversing one of three unitaries: $\hat{U}_{1}=\hat{\mathcal{I}}_{2}$, $\hat{U}_{2}$, and $\hat{U}_{3}$, where:
\begin{equation}
\hat{U}_{2}=\tfrac{1}{\sqrt{2}}\left(\begin{array}{cc}1&1\\-1&1\end{array}\right),\;\hat{U}_{3}=\tfrac{1}{\sqrt{2}}\left(\begin{array}{cc}1&-i\\-i&1\end{array}\right).
\end{equation}
After $\hat{U}_{1}=\hat{\mathcal{I}}_{2}$, $s_{0}=I_{1}+I_{2}$ and $s_{1}=I_{1}-I_{2}$; after $\hat{U}_{2}$, $s_{2}=I_{1}-I_{2}$, and after $\hat{U}_{3}$, $s_{3}=I_{1}-I_{2}$, as demonstrated recently in an on-chip platform \cite{Hashemi26arxiv}.

\subsection{Kronecker-Pauli matrices and modal Stokes parameters for a four-mode field}

We expand the $4\times4$ coherence matrix $\mathbf{G}$ for 4-mode partially coherent light in terms of $4\times4$ `Kronecker-Pauli matrices' \cite{Faddeev95IJMPA}:
\begin{equation}\label{eq:GandStokes}
\mathbf{G}=\frac{1}{4}\sum_{j,k=0}^{3}s_{jk}\hat{\sigma}_{jk},
\end{equation}
where the~16 Kronecker-Pauli matrices are outer products of Pauli matrices (Eq.~\ref{eq:Pauli}): $\hat{\sigma}_{jk}=\hat{\sigma}_{j}\otimes\hat{\sigma}_{k}$ (given explicitly in the Appendix). The modal Stokes parameters $s_{jk}$ are the real expansion coefficients of $\mathbf{G}$ in terms of $\{\hat{\sigma}_{jk}\}$, where $s_{jk}=\mathrm{Tr}\{\hat{\sigma}_{jk}\mathbf{G}\}$. This expansion has been applied to partially coherent optical fields comprising a pair of binary degrees of freedom (polarization modes and a pair of spatial modes) \cite{Abouraddy14OL,Kagalwala15SR,Harling24PRA,Harling24PRA2,Harling25APLP} (and previously to entangled photon pairs \cite{James01PRA,Abouraddy02OC}, and is applied here to four-mode partially coherent light for the first time.

The~16 Kronecker-Pauli matrices can be organized into four sets. The first set comprises diagonal matrices ($\hat{\sigma}_{00}$, $\hat{\sigma}_{01}$, $\hat{\sigma}_{10}$, and $\hat{\sigma}_{11}$), which do not require unitaries for their realization. Rather, the modal weights are directly measured: $I_{1}^{(1)}=G_{11}$, $I_{2}^{(1)}=G_{22}$, $I_{3}^{(1)}=G_{33}$, and $I_{4}^{(1)}=G_{44}$, and the associated Stokes parameters ($s_{00}$,$s_{01}$, $s_{10}$, and $s_{11}$) are obtained using Eq.~\ref{eq:S_00_01_10_11} in the Appendix [Fig.~\ref{fig:StokesParameters}(a)]. The second set of Kronecker-Pauli matrices ($\hat{\sigma}_{02}$, $\hat{\sigma}_{03}$, $\hat{\sigma}_{12}$, and $\hat{\sigma}_{13}$) are block diagonal $\left(\begin{array}{cc}\hat{A}&\hat{\mathbf{0}}_{2}\\\hat{\mathbf{0}}_{2}&\hat{B}\end{array}\right)$, with $\hat{A}$ and $\hat{B}$ corresponding to $\pm\hat{\sigma}_{2}$ and $\pm\hat{\sigma}_{3}$, and $\hat{\mathbf{0}}_{2}$ is a $2\times2$ matrix with all its elements zero. We implement unitaries $\hat{U}_{12}$ and $\hat{U}_{34}$ on the modal pairs $\{|1\rangle,|2\rangle\}$ and $\{|3\rangle,|4\rangle\}$, respectively. The modal Stokes parameters associated with $\hat{\sigma}_{02}$ and $\hat{\sigma}_{12}$ ($s_{02}$ and $s_{12}$) are evaluated by implementing $\hat{U}_{12}=\hat{U}_{34}=\hat{U}_{2}$, and then using Eq.~\ref{eq:S02_12}, whereas the modal Stokes parameters associated with $\hat{\sigma}_{03}$ and $\hat{\sigma}_{13}$ ($s_{03}$ and $s_{13}$) are evaluated by implementing $\hat{U}_{12}=\hat{U}_{34}=\hat{U}_{3}$, and then using Eq.~\ref{eq:S03_13} [Fig.~\ref{fig:StokesParameters}(b)].

The remaining~8 Kronecker-Pauli matrices have the off-diagonal block form $\left(\begin{array}{cc}\hat{\mathbf{0}}_{2}&\hat{A}\\\hat{B}&\hat{\mathbf{0}}_{2}\end{array}\right)$. One group of these ($\hat{\sigma}_{20}$, $\hat{\sigma}_{30}$, $\hat{\sigma}_{21}$, and $\hat{\sigma}_{31}$) requires implementing unitaries $\hat{U}_{13}$ and $\hat{U}_{24}$ on the modal pairs $\{|1\rangle,|3\rangle\}$ and $\{|2\rangle,|4\rangle\}$, respectively. The modal Stokes parameters associated with Kronecker-Pauli matrices $\hat{\sigma}_{20}$ and $\hat{\sigma}_{21}$ ($s_{20}$ and $s_{21}$) require setting $\hat{U}_{13}=\hat{U}_{24}=\hat{U}_{2}$ and substituting the modal weights in Eq.~\ref{eq:S20_21}, and those associated with $\hat{\sigma}_{30}$ and $\hat{\sigma}_{31}$ ($s_{30}$ and $s_{31}$) require setting $\hat{U}_{13}=\hat{U}_{24}=\hat{U}_{3}$ and then substituting the modal weights in Eq.~\ref{eq:S30_31} [Fig.~\ref{fig:StokesParameters}(c)]. Finally, for $\hat{\sigma}_{22}$, $\hat{\sigma}_{23}$, $\hat{\sigma}_{32}$, and $\hat{\sigma}_{33}$ we implement unitaries $\hat{U}_{14}$ and $\hat{U}_{23}$ on the modal pairs $\{|1\rangle,|4\rangle\}$ and $\{|2\rangle,|3\rangle\}$, respectively. The modal Stokes parameters associated with Kronecker-Pauli matrices $\hat{\sigma}_{22}$ and $\hat{\sigma}_{33}$ ($s_{22}$ and $s_{33}$) require setting $\hat{U}_{14}=\hat{U}_{23}=\hat{U}_{2}$ and substituting the modal weights in Eq.~\ref{eq:S22_33}, and those associated with $\hat{\sigma}_{23}$ and $\hat{\sigma}_{32}$ ($s_{23}$ and $s_{32}$) require setting $\hat{U}_{14}=\hat{U}_{23}=\hat{U}_{3}$ and substituting the modal weights in Eq.~\ref{eq:S23_32} [Fig.~\ref{fig:StokesParameters}(d)].

\begin{figure}[t!]
\centering
\includegraphics[width=8.6cm]{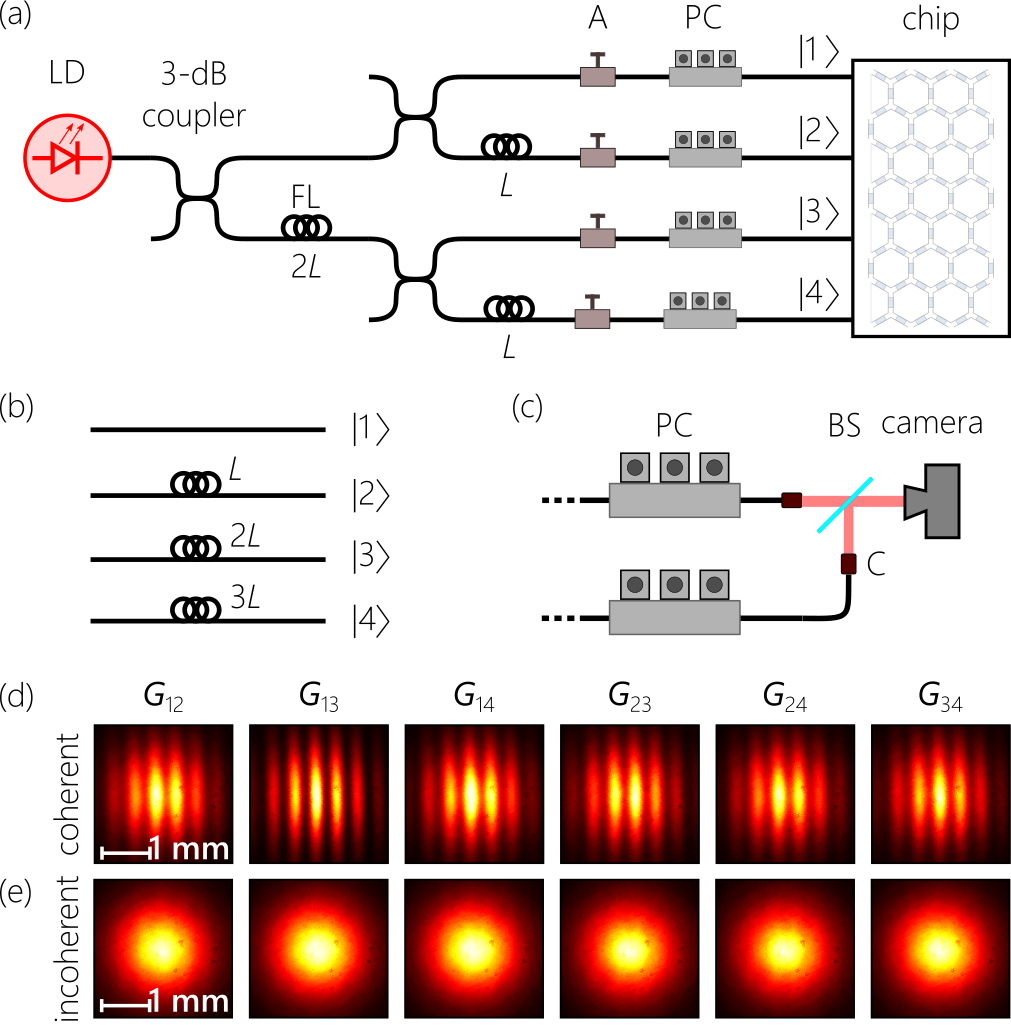} %{Fig1.jpg}
\caption{(a) Source of four-mode incoherent light. LD: Laser diode; FL: fiber loop; A: variable attenuator; PC: polarization controller. (b) The four modes $|1\rangle$, $|2\rangle$, $|3\rangle$, and $|4\rangle$ traverse relative fiber lengths of 0, $L$, $2L$, and $3L$, respectively, to reach the chip. (c) Setup for interfering a pair of modes. BS: Balanced beam splitter; C: fiber collimator. (d) Measured intensity profiles after superposing pairs of modes in absence of the fiber loops, whereupon the fields are mutually coherent and interference fringes are observed; and (e) in presence of the fiber loops, whereby the fields are mutually incoherent and the interference fringes are eliminated.}
\label{fig:Source}
\end{figure}

\section{Measurements}

\subsection{Incoherent source of four-mode light}

Our source of four-mode incoherent light described by the $4\times4$ coherence matrix $\mathbf{G}=\tfrac{1}{4}\hat{\mathcal{I}}_{4}$ (maximally incoherent light) is depicted in Fig.~\ref{fig:Source}(a). A laser diode at a wavelength $\approx1.55$~$\mu$m coupled to a single-mode fiber (SMF) is split via a 3-dB fiber coupler into paths $a$ and $b$. We place a fiber loop of length $2L$ in path $b$ (with $L\approx500$~m, which far exceeds the laser-diode coherent length). The length $L$ exceeds the fiber coherence length, so the fields in the SMFs no longer interfere at zero relative delay. We next place a 3-dB coupler in path $a$ to split it into paths $|1\rangle$ and $|2\rangle$, and a 3-dB coupler in path $b$ to split it into paths $|3\rangle$ and $|4\rangle$. We place fiber loops of length $L$ in paths $|2\rangle$ and $|4\rangle$. The lengths of fiber in paths $|1\rangle$, $|2\rangle$, $|3\rangle$, and $|4\rangle$ are $0$, $L$, $2L$, and $3L$, respectively [Fig.~\ref{fig:Source}(b)]. Attenuators placed in each path adjust the power to achieve equal values, and polarization controllers adjust the fiber polarization to match the on-chip waveguides [Fig.~\ref{fig:Source}(a)].

We thus produce four-mode incoherent light confined to four SMFs, which enables efficient coupling to on-chip single-mode waveguides. Any pair of modes are separated by at least a length $L$ [Fig.~\ref{fig:Source}(b)], so that no interference is observed when superposing the two fields. To confirm that such pairs do not exhibit mutual interference, we out-couple the fields from the SMFs through fiber collimators (F220APC-1550, Thorlabs), we overlap the fields corresponding to all six possible pairs of modes at a balanced beam splitter (BS015, Thorlabs), and record the resulting intensity distribution using a camera (Bobcat-320-GigE-13907, Xenics) placed $\approx30$~cm away from the collimators [Fig.~\ref{fig:Source}(c)]. We plot in Fig.~\ref{fig:Source}(d) the measurements after removing the fiber loops, where we observe high-visibility fringes as expected from the overlap of fields derived from the same source. In contrast, the interference fringes are eliminated in presence of the fiber loops, indicating their mutual incoherence [Fig.~\ref{fig:Source}(e)]. The visibility $V_{jk}$ for any pair of modes $|j\rangle$ and $|k\rangle$ is related to the corresponding off-diagonal element of $\mathbf{G}$, $V_{jk}=\tfrac{2|G_{jk}|}{G_{jj}+G_{kk}}$ when $\mathrm{Tr}\{\mathbf{G}\}=1$. The absence of interference for all six pairs of modes confirms that the off-diagonal elements of $\mathbf{G}$ are all zero, thus verifying that an incoherent field with $\mathbf{G}=\tfrac{1}{4}\hat{\mathcal{I}}_{4}$ is delivered to the chip.

\subsection{Controlling the coherence rank}

The first task is on-chip control of the coherence rank. Starting with rank-4 field $\mathbf{G}_{4}=\mathrm{dia}\{\tfrac{1}{4},\tfrac{1}{4},\tfrac{1}{4},\tfrac{1}{4}\}$ prepared off-chip, changing the rank requires changing the amplitudes of the diagonal elements of $\mathbf{G}$, which is a non-unitary operation realized via an MZI in the path of each mode. For example, directing mode $|4\rangle$ to one input port of an MZI [Fig.~\ref{fig:2X2Unitaries}(b), Eq.~\ref{eq:GeneralUnitary}] and providing no input to the other input port reduces the modal weight for $|4\rangle$ by $\sin^{2}(\tfrac{\delta}{2})$, discarding the other output. Setting $\delta=0$, we eliminate the mode $|4\rangle$ altogether ($\lambda_{4}=0$), thereby yielding a rank-3 field with $\mathbf{G}_{3}=\mathrm{diag}\{\tfrac{1}{3},\tfrac{1}{3},\tfrac{1}{3},0\}$. When the modal weights for both modes $|3\rangle$ and $|4\rangle$ are eliminated ($\lambda_{3}=\lambda_{4}=0$), we obtain a rank-2 field with $\mathbf{G}_{2}=\mathrm{diag}\{\tfrac{1}{2},\tfrac{1}{2},0,0\}$. Carrying out this procedure for modes $|2\rangle$, $|3\rangle$, and $|4\rangle$ ($\lambda_{2}=\lambda_{3}=\lambda_{4}=0$) yields a \textit{coherent} rank-1 field with $\mathbf{G}_{1}=\mathrm{daig}\{1,0,0,0\}$.  

\begin{figure*}[t!]
\centering
\includegraphics[width=16.5cm]{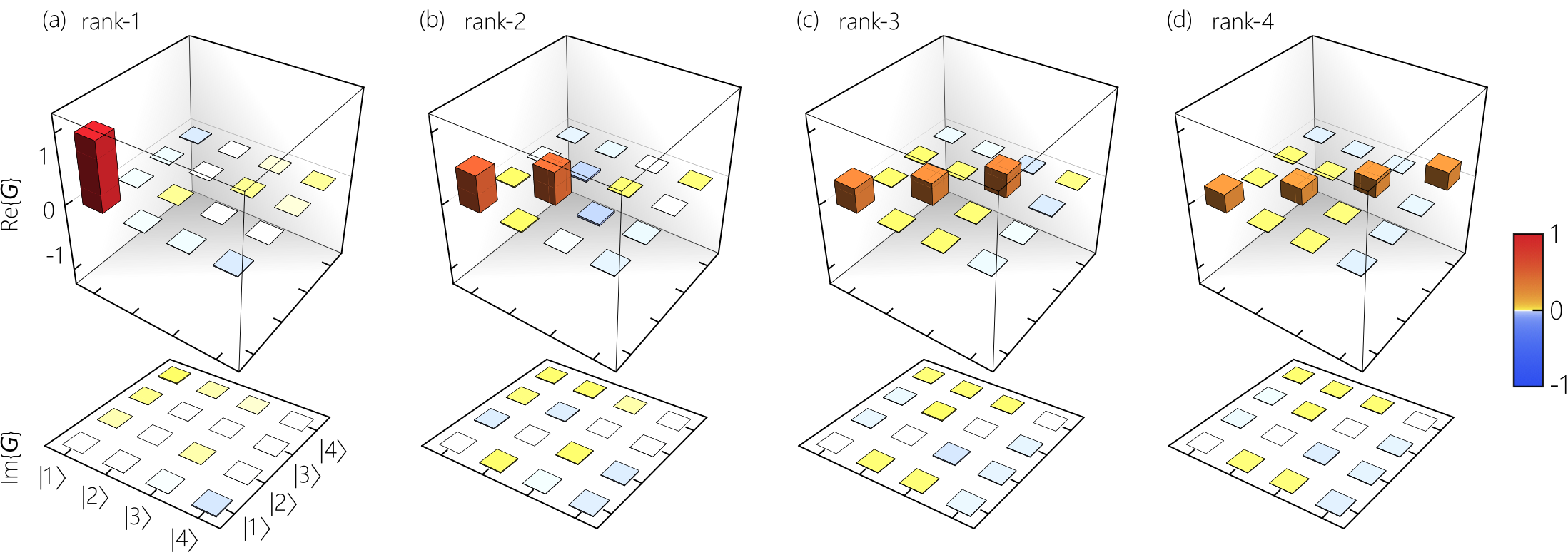} %{Fig1.jpg}
\caption{Measured coherence matrices for maximum-entropy fields of different rank. (a) Rank-1 with $\mathbf{G}_{1}=\mathrm{diag}\{1,0,0,0\}$; (b) rank-2 with $\mathbf{G}_{2}=\mathrm{diag}\{\tfrac{1}{2},\tfrac{1}{2},0,0\}$; (c) rank-3 with $\mathbf{G}_{3}=\mathrm{diag}\{\tfrac{1}{3},\tfrac{1}{3},\tfrac{1}{3},0\}$; and (d) rank-4 with $\mathbf{G}_{4}=\mathrm{diag}\{\tfrac{1}{4},\tfrac{1}{4},\tfrac{1}{4},\tfrac{1}{4}\}$. We plot separately the real and imaginary parts of the coherence matrices, $\mathrm{Re}\{\mathbf{G}\}$ and $\mathrm{Im}\{\mathbf{G}\}$, respectively.}
\label{fig:DiagonalRanks}
\end{figure*}

To reconstruct the coherence matrix $\mathbf{G}$, we program the chip to sequentially implement the four layouts depicted in Fig.~\ref{fig:StokesParameters} (right column). In each configuration the modal weights are recorded with on-chip detectors, and four~modal Stokes parameters are obtained (Appendix). The 16~modal Stokes parameters $s_{jk}$ are substituted into Eq.~\ref{eq:GandStokes} to reconstruct $\mathbf{G}$. We quantify the quality of the measured coherence matrix $\mathbf{G}_{\mathrm{meas}}$ with respect to the theoretical coherence matrix $\mathbf{G}_{\mathrm{theory}}$ using the fidelity $F=(\mathrm{Tr}\{\sqrt{\mathbf{G}_{\mathrm{meas}}}\mathbf{G}_{\mathrm{theory}}\sqrt{\mathbf{G}_{\mathrm{meas}}}\})^{2}$ \cite{Jozsa94JMO}. The measurement results for on-chip control of the coherence rank are given in Fig.~\ref{fig:DiagonalRanks}. We plot the reconstructed coherence matrices of ranks~1, 2, 3, and~4 separating the real and imaginary parts, $\mathrm{Re}\{\mathbf{G}\}$ and $\mathrm{Im}\{\mathbf{G}\}$, respectively. In all 4~cases we find negligible imaginary components, and predominantly diagonal contributions to the real part of $\mathbf{G}$. The average fidelity is $F\approx0.95$ for the reconstructed coherence matrices of rank-1, rank-2, and rank-3, and $F\approx0.99$ for rank-4. We attribute the lower fidelity to leakage of the modal amplitude from the MZIs that are utilized to extinguish the corresponding modal amplitudes, whereas the rank-4 field does not require that feature (all the modal amplitudes coupled to the chip enter the coherence-matrix reconstruction stage).

\subsection{Tuning the entropy of the coherence matrix}

On-chip tuning of the entropy $S(\mathbf{G})$ for a coherence matrix $\mathbf{G}$ of given rank requires controllably modifying the relative weights of the eigenvalues (Eq.~\ref{eq:Entropy}). We can perform this task using the same MZIs that we relied on to tune the coherence rank. However, rather than eliminating the eigenvalue altogether (when the coherence matrix is in diagonal form), we set the value of the parameter $\delta$ in Eq.~\ref{eq:GeneralUnitary} to adjust each eigenvalue by a factor $\sin^{2}(\tfrac{\delta_{j}}{2})$, where $\delta_{j}$ is the MZI parameter for the mode $|j\rangle$, $j=1,2,3,4$.

\begin{figure*}[t!]
\centering
\includegraphics[width=16.5cm]{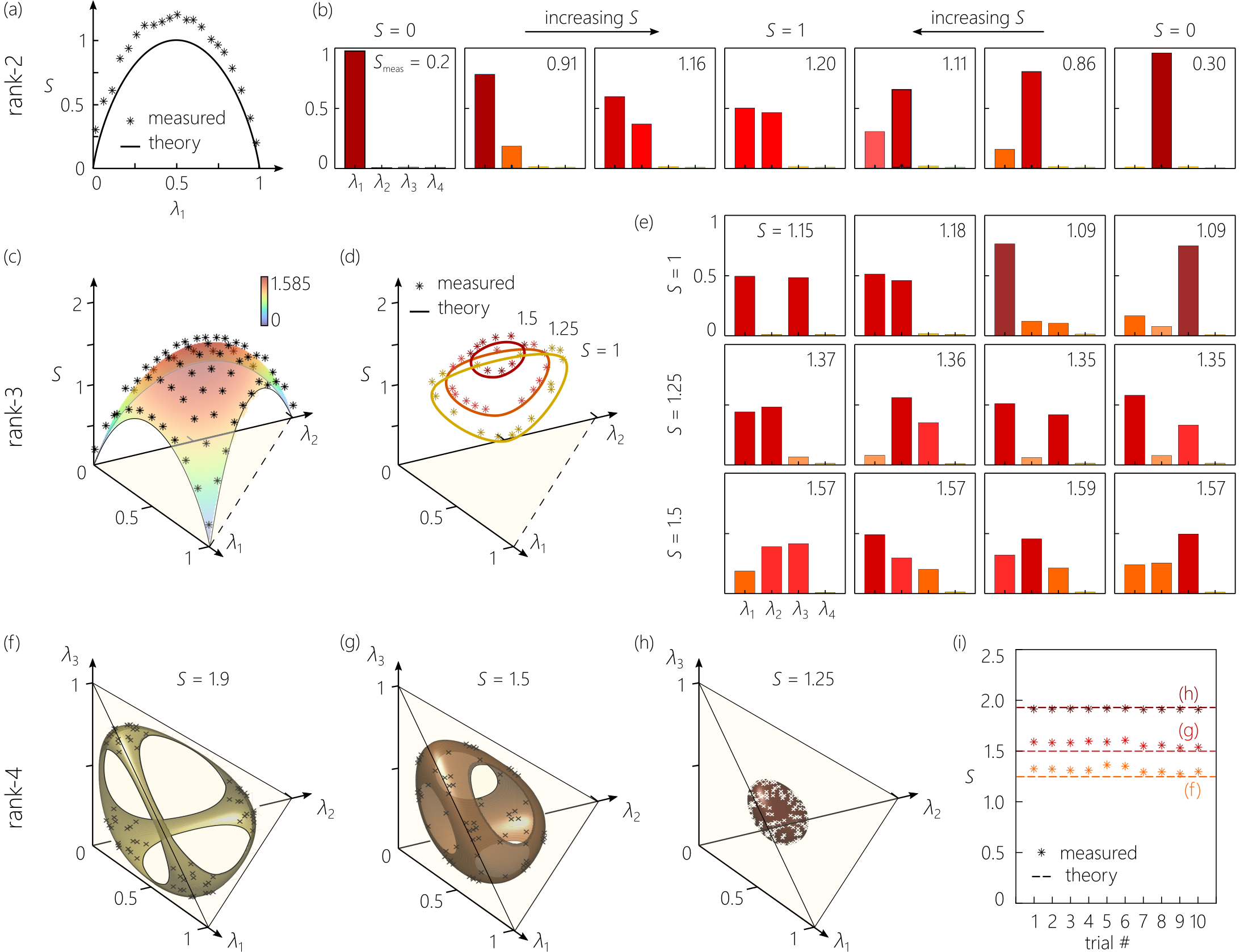} %{Fig1.jpg}
\caption{On-chip tuning of the entropy of a four-mode partially coherent field. (a) Measured entropy $S(\lambda_{1})$ for rank-2 fields as we vary the eigenvalue $\lambda_{1}$ from~0 to~1 ($\lambda_{2}=1-\lambda_{1}$, $\lambda_{3}=\lambda_{4}=0$). The curve is a theoretical plot, the points are measurements. (b) Samples of the eigenvalues of the reconstructed coherence matrices as we vary $\lambda_{1}$. The values of $S$ inside the panels are obtained from the reconstructed eigenvalues. (c) The measured entropy $S(\lambda_{1},\lambda_{2})$ for rank-3 fields as points overlaid on the theoretical surface. (d) Measured entropy for iso-entropy rank-3 fields plotted along with the theoretical curves for $S=1$, 1.25, and 1.5~bits. (e) Samples of the extracted eigenvalues from reconstructed iso-entropy coherence matrices corresponding to (d). The values of $S$ inside the panels are obtained from the reconstructed eigenvalues. (f) The measured entropy for iso-entropy rank-4 fields with $S=1.9$~bits (plotted as black points) overlaid on the theoretical iso-entropy surface. (g) Same as (f) for $S=1.5$~bits, and (h) for $S=1.25$~bits. In the latter, the measurement points are white for clarity. (i) The measured entropy for 10~reconstructed coherence matrices selected from (f-h) compared to the target entropy values.}
\label{fig:Entropy}
\end{figure*}

Rank-1 fields have $S=0$, which cannot be tuned once we set $\lambda_{2}=\lambda_{3}=\lambda_{4}=0$; changing the amplitude of the remaining modal amplitude associated with $|1\rangle$ does not change $S$. Rank-2 fields with $\mathbf{G}=\mathrm{diag}\{\lambda_{1},\lambda_{2},0,0\}$ have entropy $0<S\leq1$. We carry out measurements by eliminating $\lambda_{3}$ and $\lambda_{4}$ and then tuning the values of $\lambda_{1}$ and $\lambda_{2}$ from~0 to~1, with $\lambda_{1}+\lambda_{2}=1$. We plot the measured entropy for the reconstructed coherence matrices as points in Fig.~\ref{fig:Entropy}(a), along with the theoretical curve $S=-\lambda_{1}\log_{2}\lambda_{1}-\lambda_{2}\log_{2}\lambda_{2}$. In addition, we plot in Fig.~\ref{fig:Entropy}(b) eigenvalues $\{\lambda_{1},\lambda_{2},\lambda_{3},\lambda_{4}\}$ extracted from of a selection of reconstructed coherence matrices. We note that the measured entropy is consistently higher than the theoretical expectation. Note also that the reconstructed eigenvalues deviate only a few percent from theory; for example, in the first coherence matrix in Fig.~\ref{fig:Entropy}(b) we have $\lambda_{1}\approx0.98$ (rather than $\lambda_{1}=1$) and $\lambda_{2}+\lambda_{3}+\lambda_{4}\approx0.02$ (rather than~0). This small deviation leads to a large shift in entropy (from~0 to~$\approx0.2$~bit) because of the sensitivity of the logarithm function. The deviation is due to amplitude leakage from the MZI's used to control the relative modal weights (the amplitude is not completely extinguished as required), which necessitates furthermore improvements.

Crucially, the entropy $S$ for a rank-2 field uniquely identifies the two eigenvalues \cite{Brosseau06PO}. Any pair of rank-2 coherence matrices having equal entropy can be inter-converted unitarily. Conversely, any two coherence matrices that can be inter-converted unitarily have the same entropy \cite{Brosseau06PO}. This property does \textit{not} extend to rank-3 fields: entropy no longer identifies a coherence matrix to within a unitary \cite{Harling24PRA2}. Indeed, any rank-3 coherence matrices $\mathbf{G}_{3}$ and $\mathbf{G}_{3}'$ that can be inter-converted unitarily $\mathbf{G}_{3}'=\hat{U}\mathbf{G}_{3}\hat{U}^{\dagger}$ have the same entropy, $S(\mathbf{G}_{3})=S(\mathbf{G}_{3}')$. However, the opposite is \textit{not} guaranteed: two rank-3 coherence matrices $\mathbf{G}_{3}$ and $\mathbf{G}_{3}'$ having equal entropy $S(\mathbf{G}_{3})=S(\mathbf{G}_{3}')$ are not guaranteed to be inter-converted into each other unitarily. In such a case where the two iso-entropy rank-3 fields cannot be inter-converted unitarily, non-unitary transformations are required instead \cite{Harling24PRA2}.

We tuned the entropy of rank-3 fields ($\lambda_{4}=0$) by sweeping the values of $\lambda_{1}$, $\lambda_{2}$, and $\lambda_{3}$: $\lambda_{1}$ is swept over the range from~0 to~1, $\lambda_{2}$ is swept from~0 to $1-\lambda_{1}$, and $\lambda_{3}=1-\lambda_{1}-\lambda_{2}$. This is accomplished using MZIs associated with the modal weights of $|1\rangle$, $|2\rangle$, and $|3\rangle$. The measured entropy obtained from the reconstructed rank-3 coherence matrices are plotted as points in Fig.~\ref{fig:Entropy}(c) along with the theoretical surface $S=-\sum_{j=1}^{3}\lambda_{j}\log_{2}\lambda_{j}$. We plot the measurements and the theoretical surface in Fig.~\ref{fig:Entropy}(c) over a right-angled triangular domain in the $(\lambda_{1},\lambda_{2})$-plane delimited by the line $\lambda_{1}+\lambda_{2}=1$ (representing the rank-2 limit of rank-3 fields, $\lambda_{3}\rightarrow0$). The entropy surface reaches $S=0$ at the vertices of the triangle, corresponding to $\lambda_{1}=1$, $\lambda_{2}=1$, or $\lambda_{1}=\lambda_{2}=0$ whereupon $\lambda_{3}=1$. The entropy surface reaches a peak of $S=\log_{2}3\approx1.585$ at $\lambda_{1}=\lambda_{2}=\tfrac{1}{3}$. Any point on this surface represents all the coherence matrices that can be inter-converted into each other unitarily. All the measured values of $S$ are in close proximity to the theoretical surface.

To confirm the versatility of on-chip entropy tuning, we vary the modal weights of $|1\rangle$, $|2\rangle$, and $|3\rangle$ to produce iso-entropy fields that cannot be inter-converted unitarily. Such fields are represented in the space depicted in Fig.~\ref{fig:Entropy}(c) by the points on a curve at the intersection of the surface for $S$ with a horizontal plane corresponding to fixed entropy. The curve becomes a point when $S\rightarrow\log_{2}3$, and its area grows with decreasing $S$ to reach a maximum when $S=1$, whereupon it is tangential to the three sides of the base triangle. When $S<1$, the curve breaks up into disjoint segments in the vicinity of the vertices of the base triangle. We select $S=1,1.25,$ and 1.5~bits, and produce iso-entropy coherence matrices at these three values. The entropy obtained from the reconstructed coherence matrices are plotted as points along with the iso-entropy curves in Fig.~\ref{fig:Entropy}(d). Examples of the eigenvalues extracted from the reconstructed coherence matrices associated with these fixed entropy values are given in Fig.~\ref{fig:Entropy}(e). For each value of $S$, the iso-entropy coherence matrices have different eigenvalues and thus cannot be inter-converted unitarily.

To tune the entropy of rank-4 fields, we engage all 4~MZIs associated with the modes to control their corresponding eigenvalues. Diagonal rank-4 fields have three independent parameters after enforcing the constraints $\sum_{j=1}^{4}\lambda_{j}=1$ and $0\leq\lambda_{j}\leq1$ on the eigenvalues. A plot of the entropy for general rank-4 coherence matrices therefore cannot be readily visualized. We pursue instead iso-entropy rank-4 fields. The fixed entropy yields an additional constraint that restricts the family of rank-4 fields to a surface in $(\lambda_{1},\lambda_{2},\lambda_{3})$-space [Fig.~\ref{fig:Entropy}(f-h)]. Any rank-4 coherence matrix can be represented by a point in $(\lambda_{1},\lambda_{2},\lambda_{3})$-space within the triangular pyramid shown in Fig.~\ref{fig:Entropy}(f-h). Points on the facets correspond to rank-3 fields, points along the edges correspond to rank-2 fields, and the vertices are rank-1 fields. In Fig.~\ref{fig:Entropy}(f) we plot the theoretical iso-entropy surface corresponding to $S=1.9$~bits. Because $S=1.9>\log_{2}3$, only rank-4 fields can attain this value, and the surface lies within the volume of the triangular pyramid (no rank-2 or rank-3 fields can attain this value of entropy). The points correspond to reconstructed coherence matrices at this entropy value, which all lie within the vicinity of the theoretical surface. Measurement results for iso-entropy rank-4 fields with $S=1.5$~bits are plotted in Fig.~\ref{fig:Entropy}(g) along with the theoretical iso-entropy surface, which now intersects with the facets of the volume (rank-3 fields can attain $S=1.5$~bits, but not rank-2). The corresponding measurements for $S=1.25$~bits are plotted in Fig.~\ref{fig:Entropy}(h). Finally, the measured entropy for all the reconstructed coherence matrices in Fig.~\ref{fig:Entropy}(f-h) are collected in Fig.~\ref{fig:Entropy}(i) to facilitate comparison with the theoretical targets. In general, the data for rank-4 fields is is stronger agreement with theoretical expectations than rank-2 and rank-3 fields that require complete extinguishing of on-chip modal amplitudes.

\subsection{Unitary transformations}

\begin{figure*}[t!]
\centering
\includegraphics[width=16.5cm]{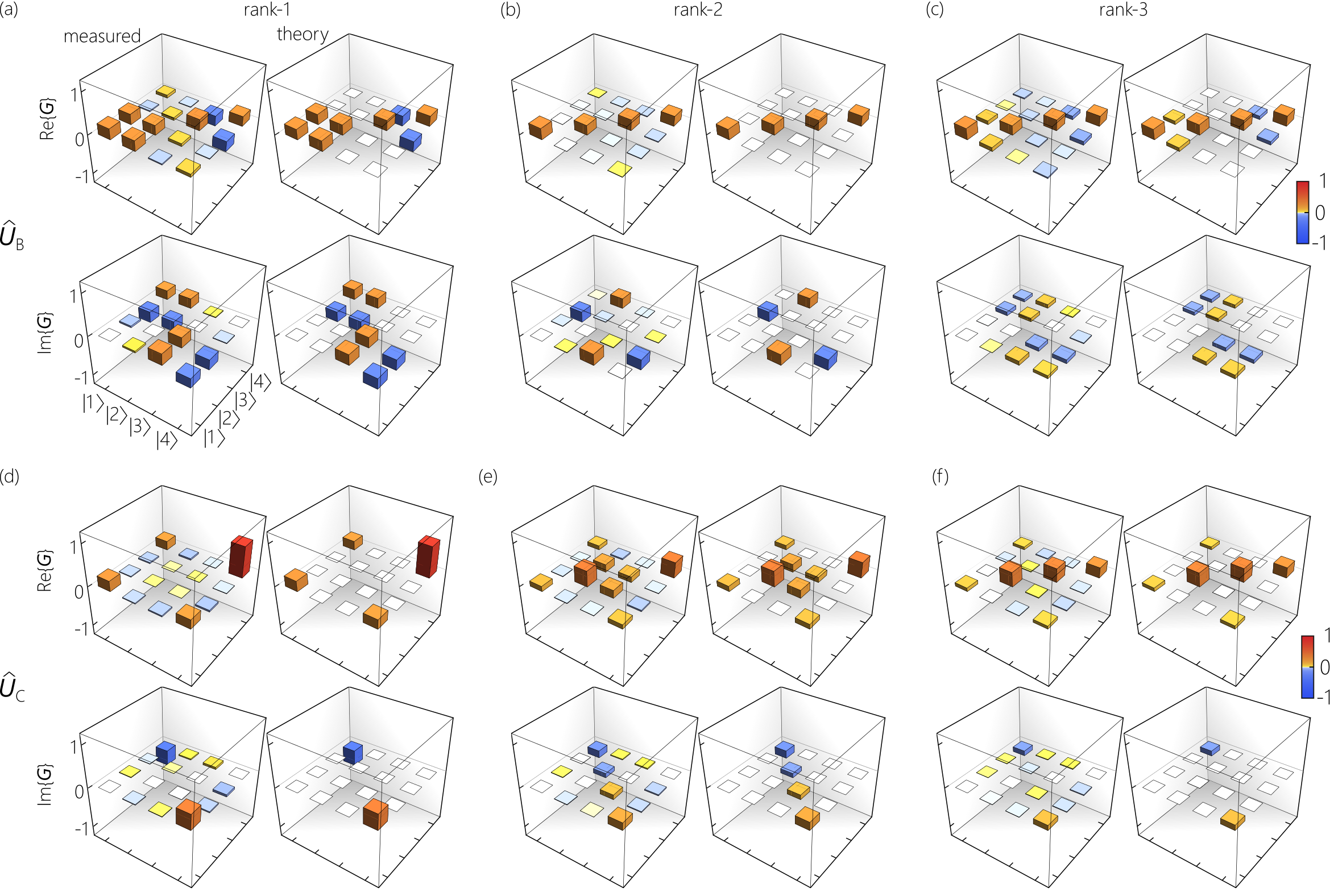} %{Fig1.jpg}
\caption{Measured $\mathbf{G}$ for rank-1, rank-2, and rank-3 fields from Fig.~\ref{fig:DiagonalRanks} after traversing the unitaries $\hat{U}_{\mathrm{B}}$ and $\hat{U}_{\mathrm{C}}$. (a) The real and imaginary parts of the rank-1 coherence matrix $\mathbf{G}=\mathrm{diag}\{1,0,0,0\}$ after traversing $\hat{U}_{\mathrm{B}}$. The left column in the panel is the measured coherence matrix, and the right column is the theoretically expected counterpart. (b) Same as (a) for the rank-2 coherence matrix $\mathbf{G}=\mathrm{diag}\{\tfrac{1}{2},\tfrac{1}{2},0,0\}$ after traversing $\hat{U}_{\mathrm{B}}$. (c) Same as (a) for the rank-3 coherence matrix $\mathbf{G}=\mathrm{diag}\{\tfrac{1}{3},\tfrac{1}{3},\tfrac{1}{3},0\}$ after traversing $\hat{U}_{\mathrm{B}}$. (d-f) Same as (a-c) for the three fields of ranks~1, 2, and 3 after traversing $\hat{U}_{\mathrm{C}}$.}
\label{fig:Rank123}
\end{figure*}

After on-chip tuning of the rank and entropy of the four-mode coherence matrix via non-unitary transformations, we proceed to modify the structure of the coherence matrix via unitary transformations. We implement the $4\times4$ unitaries $\hat{U}_{\mathrm{B}}$ and $\hat{U}_{\mathrm{C}}$ (Eq.~\ref{eq:U_BC}), whose layouts are depicted in Fig.~\ref{fig:2X2Unitaries}(f,g). We direct to these two unitaries the fields corresponding to diagonal coherence matrices, $\mathbf{G}_{1}$ through $\mathbf{G}_{4}$, corresponding to ranks~1 through~4. We plot the reconstructed coherence matrix and the theoretical expectation (separating their real and imaginary parts) after $\mathbf{G}_{1}$ traverses $\hat{U}_{\mathrm{B}}$ in Fig.~\ref{fig:Rank123}(a), after $\mathbf{G}_{2}$ traverses $\hat{U}_{\mathrm{B}}$ in Fig.~\ref{fig:Rank123}(b), and $\mathbf{G}_{3}$ traverses $\hat{U}_{\mathrm{B}}$ in Fig.~\ref{fig:Rank123}(c). The corresponding results after $\mathbf{G}_{1}$, $\mathbf{G}_{2}$, and $\mathbf{G}_{3}$ traverse $\hat{U}_{\mathrm{C}}$ in Fig.~\ref{fig:Rank123}(d-f). The unitaries were selected to mold the structure of the on-chip prepared diagonal coherence matrices and prepare new coherence matrices (having the same rank and entropy) but whose off-diagonal elements are appreciable, indicating that correlations have been introduced between the modes, so that they can now exhibit interference when superposed -- in contrast the the correlation-free state introduced into the chip as evinced in Fig.~\ref{fig:Source}(e). Moreover, the unitaries transform the coherence matrices to have significant imaginary off-diagonal values, indicating tunability of the phase of the correlation between the modes. The average fidelity of the reconstructed coherence matrices is $F\approx0.95$. For the rank-4 field $\mathbf{G}_{4}=\tfrac{1}{4}\hat{\mathcal{I}}_{4}$, the unitaries $\hat{U}_{\mathrm{B}}$ and $\hat{U}_{\mathrm{C}}$ leave the coherence matrix invariant as expected, $\hat{U}\mathbf{G}_{4}\hat{U}^{\dagger}=\mathbf{G}_{4}$. In all cases, the fidelity $F$ exceeds $\approx0.95$. These results confirm the feasibility of modifying the structure of $4\times4$ coherence matrices via unitaries in an integrated photonics platform. 

\section{Discussion and conclusion}

Our results here provide the first experimental confirmation for the scalability of on-chip manipulation of multimode partially coherent light using basic building blocks that transform pairs of modes at a time. This conclusion with regards to scalability extends to all the facets necessary for full control over partially coherent light in an integrated-photonics platform with intended utility in optical information processing. First, our approach for the efficient coupling of partially coherent light to single-mode on-chip waveguides can be readily expanded to $N\geq4$~modes by using a larger number of fiber couplers and delays, with the potential addition of fiber amplifiers if needed. Second, on-chip control of the coherence rank requires only a single element (an MZI or a switch) per mode to selectively eliminate the corresponding modal weight. Third, the field entropy can be tuned on chip using a single MZI per mode to precisely modify the corresponding modal weight. Both of these tasks, controlling the coherence rank and tuning the field entropy, are linear non-unitary operations that can be readily extended to $N\geq4$~modes. Fourth, the structure of the $N\times N$ coherence matrix can be molded on chip through concatenation of $2\times2$ unitary building blocks \cite{Reck94PRL,Pai19PRA,Bell21APLP} -- as long as a sufficient number of on-chip building blocks are available -- utilizing the approach that has been well-established in mathematics, quantum optics, and programmable photonics \cite{Bogaerts20Nature}. Fifth, tomographic reconstruction of the $N\times N$ Hermitian coherence matrix can be extended to $N\geq4$ utilizing Kronecker-Pauli matrices of corresponding dimension, which in turn determine the requisite measurements for the modal Stokes parameters.

Several open questions need to be addressed to further the utility of programmable photonics for manipulating partially coherent light. First, for precise tuning of the field entropy it is necessary to eliminate any leakage from the MZIs, which refers to a minute unwanted non-zero transmission through an output port when a zero output is targeted. Second, whereas the reconstruction of $N\times N$ coherence matrices is straightforward for $N$~even, odd-dimensional modal basis requires adopting a different matrix decomposition of the coherence matrix, likely by utilizing Gell-Mann matrices, which generate the elements of the unitary matrix group SU(3) \cite{Griffiths08BookElementaryParticles}, as the Pauli matrices generate the elements of SU(2). Third, as the dimension $N$ increases, the speed of coherence-matrix reconstruction will become a bottleneck. This requires a careful comparison of the modal-Stokes-parameters reconstruction strategy presented here (see also \cite{Kagalwala13NP,Abouraddy14OL,Kagalwala15SR}, which requires a fixed number of $N^{2}$ measurements carried out on any $N\times N$ coherence matrix, and the variational strategy outlined theoretically in Ref.~\cite{roques2024Light}, possibly augmented with machine-learning algorithms for speed-up \cite{Pereira26arxiv}. 

Our findings pave the way to the preparation of multimode structured coherence to be launched from integrated-photonics platforms into free space or multimode fibers and exploit the coherence advantage in optical \cite{Nardi22OL,Harling25APLP} or millimeter wave \cite{Sanjari23NC} communications. The results reported here validate the utility of programmable photonics in manipulating partially coherent light for fundamental studies of entropy concentration and swapping \cite{Okoro17Optica,Harling22OE,Harling24PRA,Harling24PRA2}, and in high-speed applications in secure optical communications \cite{Wacogne96IEEEPTL,Rhodes16AO}, sensing \cite{Baleine04OL,Redding12NP}, cryptography \cite{Peng21P,Liu25LPR}, computation \cite{Dong24Nature}, and spectroscopy \cite{Miller25Optica}.

%\section{Conclusions}

%In conclusion, we have demonstrated on-chip control over four-mode partially coherent optical fields. Utilizing an integrated hexagonal mesh of MZIs, we carry out the fundamental tasks needed to exploit partial coherence on an optical chip: (1) controlling the coherence rank; (2) tuning the entropy of the coherence matrix (both of which correspond to non-unitary transformations); (3) molding the coherence matrix via $4\times4$ unitary transformations); and (4) reconstructing the $4\times4$ spatial coherence matrix by measuring the spatial Stokes parameters, which are the expansion coefficients of the Hermitian coherence matrix in terms of composite Pauli matrices. These tasks are preceded by efficiently coupling to the chip generic four-mode spatially incoherent light, loaded from single-mode fibers to single-mode on-chip waveguides.

\section*{Appendix: Measuring the modal Stokes parameters og four-mode partially coherent light}

The Kronecker-Pauli matrices $\{\hat{\sigma}_{jk}\}$ are direct products of Pauli matrices. The Kronecker-Pauli matrices $\hat{\sigma}_{00}$, $\hat{\sigma}_{01}$, $\hat{\sigma}_{10}$, and $\hat{\sigma}_{11}$ are: 
\begin{eqnarray}\label{eq:Sigma00_01_10_11}
\hat{\sigma}_{00}&=&\left(\begin{array}{cccc}1&0&0&0\\0&1&0&0\\0&0&1&0\\0&0&0&1\end{array}\right),
\hat{\sigma}_{01}=\left(\begin{array}{cccc}1&0&0&0\\0&-1&0&0\\0&0&1&0\\0&0&0&-1\end{array}\right),\nonumber\\
\hat{\sigma}_{10}&=&\left(\begin{array}{cccc}1&0&0&0\\0&1&0&0\\0&0&-1&0\\0&0&0&-1\end{array}\right),
\hat{\sigma}_{11}=\left(\begin{array}{cccc}1&0&0&0\\0&-1&0&0\\0&0&-1&0\\0&0&0&1\end{array}\right),
\end{eqnarray}
which are diagonal, so that they correspond to direct measurements of the modal weights $I_{j}^{(1)}$ [Fig.~\ref{fig:StokesParameters}(a)], from which the associated modal Stokes parameters $s_{00}$, $s_{01}$, $s_{10}$, and $s_{11}$ can be evaluated as follows:
\begin{eqnarray}\label{eq:S_00_01_10_11}
s_{00}&=&I_{1}^{(1)}+I_{2}^{(1)}+I_{3}^{(1)}+I_{4}^{(1)},\nonumber\\
s_{01}&=&I_{1}^{(1)}-I_{2}^{(1)}+I_{3}^{(1)}-I_{4}^{(1)},\nonumber\\
s_{10}&=&I_{1}^{(1)}+I_{2}^{(1)}-I_{3}^{(1)}-I_{4}^{(1)},\nonumber\\
s_{11}&=&I_{1}^{(1)}-I_{2}^{(1)}-I_{3}^{(1)}+I_{4}^{(1)}.
\end{eqnarray}

The Kronecker-Pauli matrices $\hat{\sigma}_{02}$, $\hat{\sigma}_{03}$, $\hat{\sigma}_{12}$, and $\hat{\sigma}_{13}$ are:
\begin{eqnarray}\label{eq:Sigma02_03_12_13}
\hat{\sigma}_{02}&=&\left(\begin{array}{cccc}0&1&0&0\\1&0&0&0\\0&0&0&1\\0&0&1&0\end{array}\right),
\hat{\sigma}_{03}=\left(\begin{array}{cccc}0&-i&0&0\\i&0&0&0\\0&0&0&-i\\0&0&i&0\end{array}\right),\nonumber\\
\hat{\sigma}_{12}&=&\left(\begin{array}{cccc}0&1&0&0\\1&0&0&0\\0&0&0&-1\\0&0&-1&0\end{array}\right),
\hat{\sigma}_{13}=\left(\begin{array}{cccc}0&-i&0&0\\i&0&0&0\\0&0&0&i\\0&0&-i&0\end{array}\right),
\end{eqnarray}
which have the block diagonal form $\left(\begin{array}{cc}\hat{A}&\hat{\mathbf{0}}_{2}\\\hat{\mathbf{0}}_{2}&\hat{B}\end{array}\right)$, with $\hat{A}$ and $\hat{B}$ corresponding to $\pm\hat{\sigma}_{2}$ and $\pm\hat{\sigma}_{3}$. The associated modal Stokes parameters $s_{02}$, $s_{03}$, $s_{12}$, and $s_{13}$ are obtained by implementing unitaries $\hat{U}_{12}$ and $\hat{U}_{34}$ and recording the modal weights $I_{j}^{(2)}$; setting $\hat{U}_{12}=\hat{U}_{34}=\hat{U}_{2}$ yields:
\begin{eqnarray}\label{eq:S02_12}
s_{02}&=&I_{1}^{(2)}-I_{2}^{(2)}+I_{3}^{(2)}-I_{4}^{(2)},\nonumber\\
s_{12}&=&I_{1}^{(2)}-I_{2}^{(2)}-I_{3}^{(2)}+I_{4}^{(2)},
\end{eqnarray}
and setting $\hat{U}_{12}=\hat{U}_{34}=\hat{U}_{3}$ yields:
\begin{eqnarray}\label{eq:S03_13}
s_{03}&=&I_{1}^{(2)}-I_{2}^{(2)}+I_{3}^{(2)}-I_{4}^{(2)},\nonumber\\
s_{13}&=&I_{1}^{(2)}-I_{2}^{(2)}-I_{3}^{(2)}+I_{4}^{(2)}.
\end{eqnarray}

The Kronecker-Pauli matrices $\hat{\sigma}_{20}$, $\hat{\sigma}_{21}$, $\hat{\sigma}_{30}$, and $\hat{\sigma}_{31}$ are:
\begin{eqnarray}\label{eq:Sigma20_21_30_31}
\hat{\sigma}_{20}&=&\left(\begin{array}{cccc}0&0&1&0\\0&0&0&1\\1&0&0&0\\0&1&0&0\end{array}\right),
\hat{\sigma}_{21}=\left(\begin{array}{cccc}0&0&1&0\\0&0&0&-1\\1&0&0&0\\0&-1&0&0\end{array}\right),\nonumber\\
\hat{\sigma}_{30}&=&\left(\begin{array}{cccc}0&0&-i&0\\0&0&0&-i\\i&0&0&0\\0&i&0&0\end{array}\right),
\hat{\sigma}_{31}=\left(\begin{array}{cccc}0&0&-i&0\\0&0&0&i\\i&0&0&0\\0&-i&0&0\end{array}\right),
\end{eqnarray}
which have the off-diagonal block-matrix form $\left(\begin{array}{cc}\hat{\mathbf{0}}_{2}&\hat{A}\\\hat{B}&\hat{\mathbf{0}}_{2}\end{array}\right)$. To measure the associated modal Stokes parameters $s_{20}$ and $s_{21}$ we implement unitaries $\hat{U}_{13}$ and $\hat{U}_{24}$ and record the modal weights $I_{j}^{(3)}$ [Fig.~\ref{fig:StokesParameters}(c)]; setting $\hat{U}_{13}=\hat{U}_{24}=\hat{U}_{2}$ yields:
\begin{eqnarray}\label{eq:S20_21}
s_{20}&=&I_{1}^{(3)}+I_{2}^{(3)}-I_{3}^{(3)}-I_{4}^{(3)},\nonumber\\
s_{21}&=&I_{1}^{(3)}-I_{2}^{(3)}-I_{3}^{(3)}+I_{4}^{(3)};
\end{eqnarray}
and setting $\hat{U}_{13}=\hat{U}_{24}=\hat{U}_{3}$ yields:
\begin{eqnarray}\label{eq:S30_31}
s_{30}&=&I_{1}^{(3)}+I_{2}^{(3)}-I_{3}^{(3)}-I_{4}^{(3)},\nonumber\\
s_{31}&=&I_{1}^{(3)}-I_{2}^{(3)}-I_{3}^{(3)}+I_{4}^{(3)}.
\end{eqnarray}

Finally, the Kronecker-Pauli matrices $\hat{\sigma}_{22}$, $\hat{\sigma}_{23}$, $\hat{\sigma}_{32}$, and $\hat{\sigma}_{33}$ are:
\begin{eqnarray}\label{eq:Sigma22_23_32_33}
\hat{\sigma}_{22}&=&\left(\begin{array}{cccc}0&0&0&1\\0&0&1&0\\0&1&0&0\\1&0&0&0\end{array}\right),
\hat{\sigma}_{23}=\left(\begin{array}{cccc}0&0&0&-i\\0&0&i&0\\0&-i&0&0\\i&0&0&0\end{array}\right),\nonumber\\
\hat{\sigma}_{32}&=&\left(\begin{array}{cccc}0&0&0&-i\\0&0&-i&0\\0&i&0&0\\i&0&0&0\end{array}\right),
\hat{\sigma}_{33}=\left(\begin{array}{cccc}0&0&0&-1\\0&0&1&0\\0&1&0&0\\-1&0&0&0\end{array}\right),
\end{eqnarray}
which again have off-diagonal block-matrix form. The associated modal Stokes parameters $s_{22}$, $s_{23}$, $s_{32}$, and $s_{33}$ require implementing unitaries $\hat{U}_{14}$ and $\hat{U}_{23}$ and recording the modal weights $I_{j}^{(4)}$ [Fig.~\ref{fig:StokesParameters}(d)]; setting $\hat{U}_{14}=\hat{U}_{23}=\hat{U}_{2}$ yields:
\begin{eqnarray}\label{eq:S22_33}
s_{22}&=&I_{1}^{(4)}+I_{2}^{(4)}-I_{3}^{(4)}-I_{4}^{(4)},\nonumber\\
s_{33}&=&-I_{1}^{(4)}+I_{2}^{(4)}-I_{3}^{(4)}+I_{4}^{(4)},
\end{eqnarray}
and setting $\hat{U}_{14}=\hat{U}_{23}=\hat{U}_{3}$ yields:
\begin{eqnarray}\label{eq:S23_32}
s_{23}&=&I_{1}^{(4)}-I_{2}^{(4)}+I_{3}^{(4)}-I_{4}^{(4)},\nonumber\\
s_{32}&=&I_{1}^{(4)}+I_{2}^{(4)}-I_{3}^{(4)}-I_{4}^{(4)}.
\end{eqnarray}

Once the modal Stokes parameters are obtained, the elements 16~elements $G_{jk}$ of the $4\times$ coherence matrix $\mathbf{G}$ can be determined:
\begin{eqnarray}
G_{11}&=&s_{00}+s_{01}+s_{10}+s_{11},\nonumber\\
G_{12}&=&s_{02}+s_{12}-i(s_{03}+s_{13}),\nonumber\\
G_{13}&=&s_{20}+s_{21}-i(s_{30}+s_{31}),\nonumber\\
G_{14}&=&s_{22}-s_{33}-i(s_{23}+s_{32}),\nonumber\\
G_{21}&=&s_{02}+s_{12}+i(s_{03}+s_{13}),\nonumber\\
G_{22}&=&s_{00}-s_{01}+s_{10}-s_{11},\nonumber\\
G_{23}&=&s_{22}+s_{33}+i(s_{23}-s_{32}),\nonumber\\
G_{24}&=&s_{20}-s_{21}-i(s_{30}-s_{31}),\nonumber\\
G_{31}&=&s_{20}+s_{21}+i(s_{30}+s_{31}),\nonumber\\
G_{32}&=&s_{22}+s_{33}-i(s_{23}-s_{32}),\nonumber\\
G_{33}&=&s_{00}+s_{01}-s_{01}-s_{11},\nonumber\\
G_{34}&=&s_{02}-s_{12}-i(s_{03}-s_{13}),\nonumber\\
G_{41}&=&s_{22}-s_{33}+i(s_{23}+s_{32}),\nonumber\\
G_{42}&=&s_{20}-s_{21}+i(s_{30}-s_{31}),\nonumber\\
G_{43}&=&s_{02}-s_{12}+i(s_{03}-s_{13}),\nonumber\\
G_{44}&=&s_{00}-s_{01}-s_{10}+s_{11}.
\end{eqnarray}
\section*{Acknowledment}
We thank Guifang Li for lending equipment.
\section*{Funding Sources}
This work was supported by the U.S. Office of Naval Research (ONR) under contract N00014-20-1-2789.
\bibliography{diffraction.bib}
\end{document}